\documentclass[10pt,letterpaper]{article}

\usepackage{amsmath,amssymb}  
\usepackage{textcomp} 
\usepackage{graphicx}
\usepackage{amsmath}
\usepackage{amssymb}
\usepackage{opex3}
\newcommand{\uM}{~$\mu$M\,}

\begin{document}

\title{Femtosecond lasing from a fluorescent protein in a one dimensional random cavity}

\author{T.M.~Drane$^{1,4}$, H.~Bach$^3$, M.~Shapiro$^{1,4}$, and V.~ Milner$^{2,4,*}$}

\address{Departments of $^1$Chemistry, $^2$Physics \& Astronomy and $^3$Medicine and $^4$The Laboratory for Advanced Spectroscopy and Imaging Research (LASIR), The University of British Columbia, Vancouver, Canada}

\email{$^*$vmilner@phas.ubc.ca}

\begin{abstract}
We present evidence of ultrafast random lasing from the fluorescent protein DsRed2 embedded in a random one-dimensional cavity. Lasing is achieved when a purified protein solution, placed inside a layered random medium, is optically excited with a femtosecond pump pulse in the direction perpendicular to the plane of random layers. We demonstrate that pumping with ultrashort pulses resulted in a lasing threshold two orders of magnitude lower than that found for nanosecond excitation.
\end{abstract}



\section{Introduction}
Bio-materials present an exciting new platform for coherent light generation and amplification. Pioneering work by Pikas \textit{et al.} demonstrated lasing in an optically excited neat solution of wild type green fluorescent protein (\textit{wt}-GFP)\cite{Pikas2002} and recent work by Gather and Yun has shown the feasibility of using GFP expressed in mammalian\cite{Gather2011} or bacterial\cite{Gather2011a} cells as a laser gain medium, in some cases achieving sub-nanojoule lasing thresholds or optical gain at chromophore concentrations as low as 2.5\uM. To date, each observation of protein lasing has required a conventional, high finesse optical cavity to provide the resonant feedback necessary for the lasing process. Since high-quality alignment-sensitive arrangements of optical components are unlikely to be realized in biological settings, the question arises as to whether lasing can be supported by living bio-systems. Such biological lasers may prove useful as probes of biological microstructure, light sources for micro spectroscopy, or ``guide-stars'' for aberration correction in tissue.

So called ``random lasers'' eschew typical resonator design and instead rely on multiple scattering events in a disordered medium to provide coherent optical feedback in an amplifying material (for a recent review of the topic, see Ref.\cite{Wiersma2008}). Random scattering of photons, which often dominates light propagation in biological materials that do not otherwise exhibit strong absorption, degrades both the spatial and temporal coherence of light, and is therefore considered a hindrance to most optical experiments. At the same time, the increased amount of scattering causes a photon to spend longer time inside the scattering material. In the presence of an amplifying medium, long photon dwell time leads to the increased probability of stimulated emission and the possibility of lasing action\cite{Cao2000}.

In essence, the scattering medium plays the role of a random laser cavity. Typically the low quality factor of such cavities results in a high pump power required to initiate the lasing, often too high for biological media to withstand. Documented random lasing in several biological media\cite{Smuk2011,Polson2004,Song2010} has been achieved only by using high-gain non-biological chromophores, such as laser dyes. The lasing threshold can be lowered to avoid optical damage through the use of high gain materials, but the toxicity of these reduces the value of random lasing for biological and medical applications.

In this work we present evidence of random lasing which is achieved both with a low-power pump source and a biologically friendly gain medium, the red fluorescent protein DsRed2\cite{Matz1999,Gross2000}. We place a concentrated solution of purified DsRed2 inside a one-dimensional (1D) random layered structure which represents one of the simplest highly scattering geometries. Periodic layered structures are well known as photonic bandgap materials. However, when the layer thicknesses are governed by a random, instead of periodic, distribution, the bandgap disappears and the transmission spectrum appears as a random distribution of peaks with varying line widths and mode spacing\cite{Chabanov2000}. By comparing the nanosecond and femtosecond regimes of excitation, we show that ultrashort pump pulses offer substantially lower threshold values, down to the level which may prove tolerable for living biological systems.
\begin{figure}[tb]
   \centering
   \includegraphics[width=.85\columnwidth]{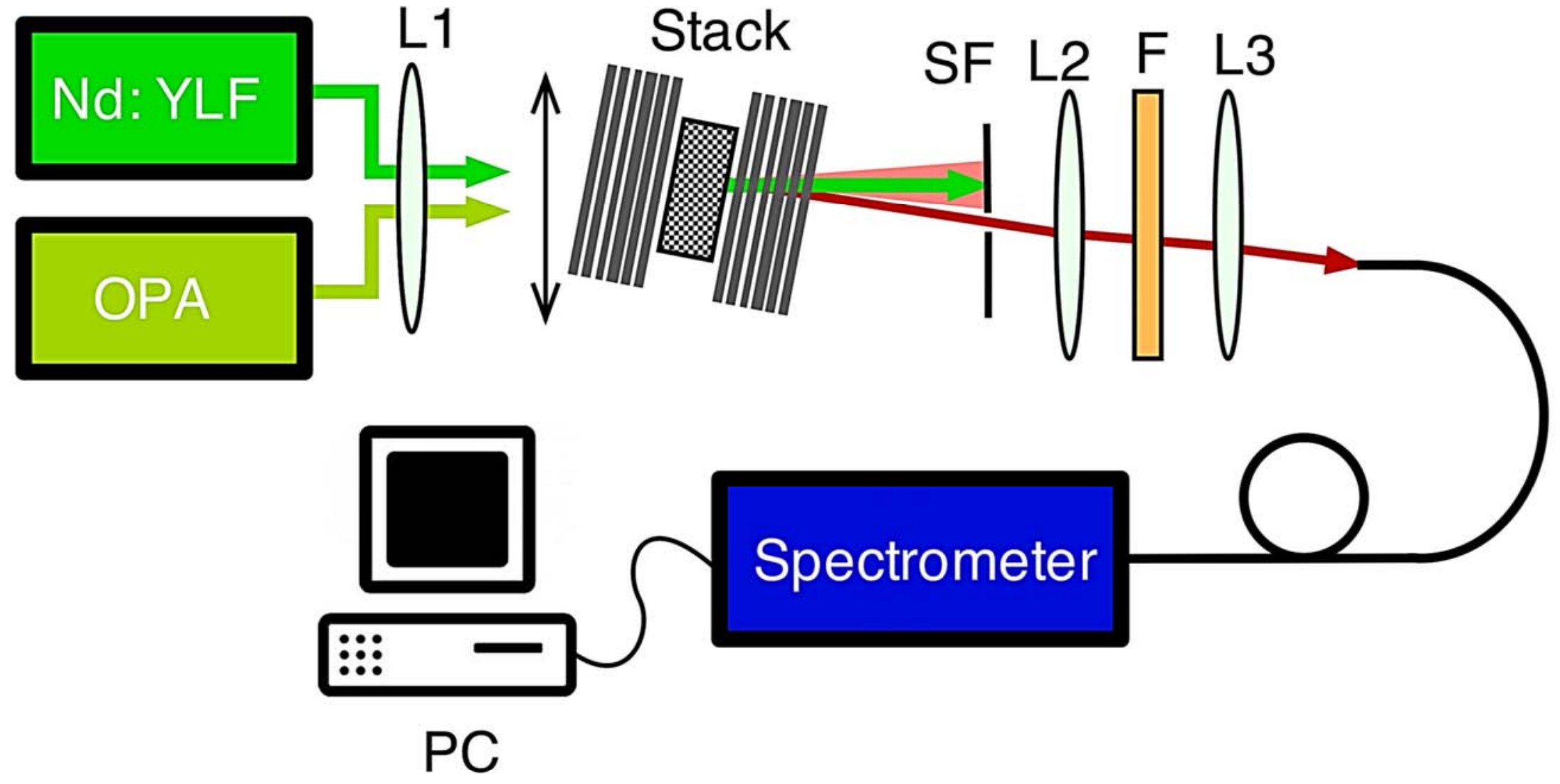}
   \caption{Experimental setup. Nd:YLF, nanosecond source; OPA, femtosecond source; L1, 400 mm focussing lens; Stack, glass coverslip and protein medium assembly mounted on translation stage; SF, iris; L2, L3 collimation \& fiber coupling lenses; F, dichroic mirror. The stack assembly is tilted relative to the pump beam axis to spatially separate the resonant lasing output from the unabsorbed pump beam and any amplified spontaneous emission. The stack was scanned in the plane transverse to the pump beam in order to illuminate different cavity configurations arising from the inhomogeneity of both the coverslip and the air gap thickness.}
   \label{Figure1}
\end{figure}

\section{Experimental setup}
The experimental setup is illustrated in Figure \ref{Figure1}. A 1D scatterer may be realized by assembling a structure of alternating layers from two transparent dielectric materials with different refractive indices (hereafter referred to as a ``random stack''). In this work, similarly to Ref.~\cite{Milner2005}, we used microscope coverslips of random thickness between 85 and 130 microns separated by air gaps. A beam of light at normal incidence to the layers of this structure is reflected at each air-glass interface. The inherent slide-to-slide variation of glass thickness is much larger than the optical wavelength, which results in an a complete phase scrambling of the multiply scattered light waves. The scattering can be considered 1D as long as the layer thickness is uniform on the scale of the transverse beam divergence, which limits the total number of glass layers to about 60\cite{Zhang2008}.

Aside from the geometrical simplicity, 1D scattering systems are particularly suitable for low-threshold random lasing because of the effect of localization of light. Localization, which is easier to achieve in the systems of lower dimensionality\cite{berry:1997}, further extends the photon residence time inside the scattering material\cite{sheng:1990}, resulting in correspondingly higher gain in the presence of an amplifying medium. Localization-based lasing has been discussed extensively in theoretical\cite{Burin2002,Chang2003} and experimental\cite{Milner2005,Monguzzi2010a} works.

As a gain medium, we used the purified and concentrated solution of the red fluorescent protein DsRed2 (see Methods for further details). The final DsRed2 protein concentration used for this study was 600 \uM. In order to introduce the protein solution inside the scattering structure, a 4 mm spacer with an 8 mm diameter through hole has been placed in the middle of the random stack and filled with the solution. To further investigate the dependence of the lasing threshold on the pumping time scale in this structure, the protein medium was replaced by the laser dyes, either Rhodamine B (for threshold characterization and fluorescence lifetime dependence) or LDS 750 (for fluorescence lifetime dependence).

The assembly was pumped by either the second harmonic from a Nd:YLF diode pumped solid state laser (529 nm, ~300 ns pulse width, Newport Empower) or by the output of a femtosecond optical parametric amplifier (OPA, 530 nm central wavelength, 35 fs pulse width, Light Conversion TOPAS-C). Pumping was performed at a small angle to the stacking axis, i.e. the axis perpendicular to the layers of the random stack (Fig.\ref{Figure1}).This geometry enabled an easy spatial separation between the amplified spontaneous emission (ASE) observed along the pumping direction, and the random lasing in the direction of the stacking axis. Each pump beam was focused onto the stack by a 400 mm lens (L1), resulting in a nominal diameter in the protein solution (measured without the stack in place) of 350~$\mu$m for the nanosecond and 245~$\mu$m for the femtosecond source, respectively.

The entire random stack assembly was mounted on a 2D translation stage which allowed us to scan the location of the pump beam on the stack face without altering the pump pointing. All reported pump powers were measured immediately before the stack. The forward propagating photoluminescence was relayed by a pair of 100~mm lenses (L2, L3) into a multimode fiber coupled to a f/4.8, 0.05~nm resolution spectrometer (McPhearson 2035, femtosecond pumping).  A dichroic filter (F), long pass at 580~nm, placed before the fiber input coupler removed any unabsorbed pump energy. An iris (Fig.\ref{Figure1}, SF) placed before L2 blocks the residual pump beam as well as any ASE generated.  To prevent immediate bleaching of the pumped area and to allow single-shot data collection of the emission spectra, the femtosecond pump was operated at 20 Hz while a mechanical chopper was used to reduce the repetition rate of the nanosecond excitation to 20 Hz as well.

\section{Results}
\subsection{Line narrowing}
\begin{figure*}[tb]
\centering
\includegraphics[width=.85\columnwidth]{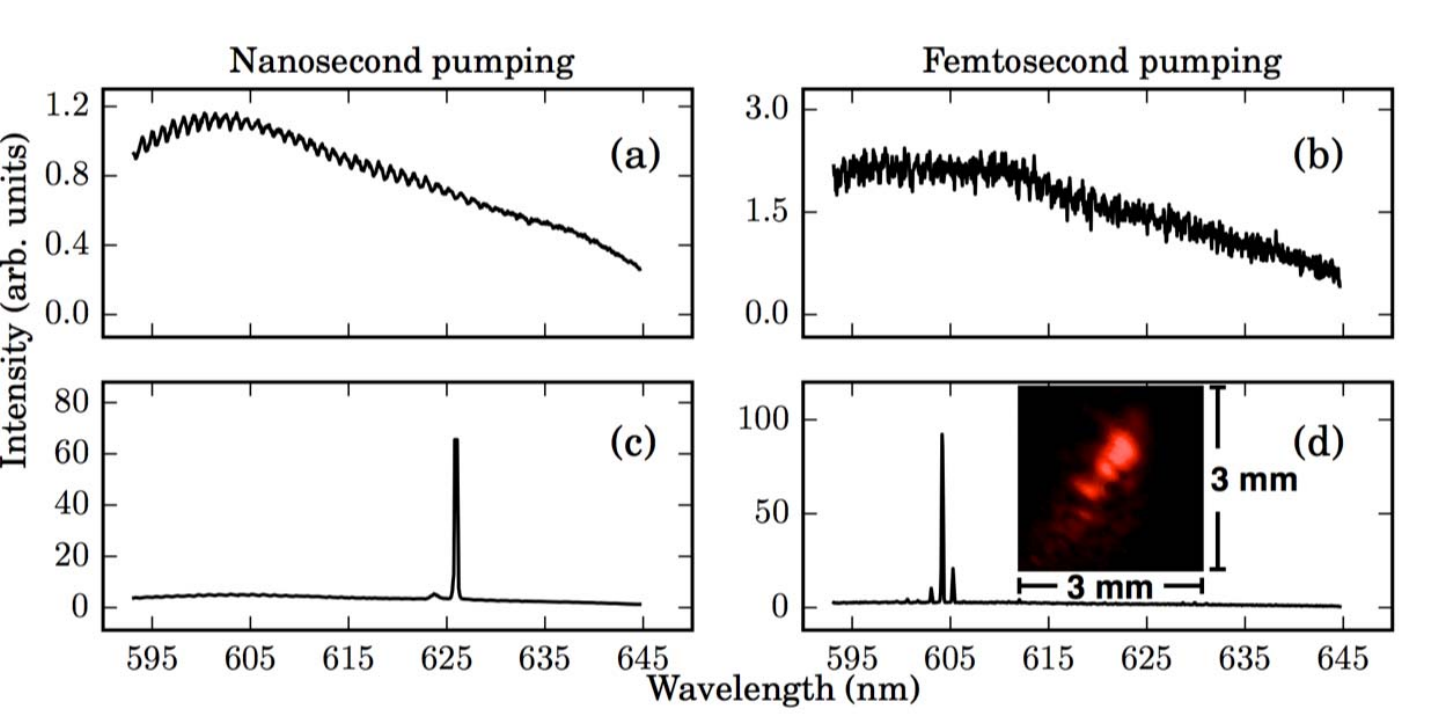}
\caption{Typical spectra of DsRed emission under nanosecond (a, c) and femtosecond (b, d) pumping. Below the lasing threshold (a, b) the  emission spectrum is broad ($>$ 40~nm FWHM) and shows the quasi-periodic modulation resulting from the $\sim$100~$\mu$m coverslip thickness. Above the lasing threshold (c, d) narrow lasing modes appear above the spontaneous emission background. Inset (d): Profile of the emitted beam approximately 5 cm from the output face of the random stack.}
\label{Figure2}
\end{figure*}

We observed random lasing from DsRed under both nanosecond and femtosecond excitation, with ultrafast excitation exhibiting significantly different threshold power. In both cases, the transition to lasing was identified by a sudden narrowing of the emission spectrum when the pump power was increased above threshold. At lower-than-threshold pump power, the spectrum of the emitted light is as broad as that of the DsRed fluorescence which spans more than 50 nm, as shown in Fig.\ref{Figure2}(a, b). The broad fluorescence line shape is modulated by a quasi-periodic transmission spectrum of the random stack \cite{DraneJMO} as shown in \ref{Figure2}(a).

When the energy of the pump pulses exceeded the threshold value, the spectrum of the emitted light showed a number of intense, distinct peaks distributed randomly throughout the gain bandwidth. This is shown in Figures \ref{Figure2} (c) and (d) for the cases of nanosecond and femtosecond pumping, respectively. These spectral features had a line width below 0.1 nm, limited by the resolution of our spectrometer, and are significantly narrower than the width of the protein fluorescence spectrum.

Above the lasing threshold a lasing beam visible to the unaided eye (\ref{Figure2}(d, inset)) was emitted from the stack in both the forward and backward direction regardless of the pump pulse time scale. Aside from the significantly lower threshold figures, discussed below, the lasing output characteristics from femtosecond pumping were otherwise indistinguishable from those found with nanosecond pumping.

\subsection{Lasing threshold}
\begin{figure}[tb]
\centering
 \includegraphics[width=.85\columnwidth]{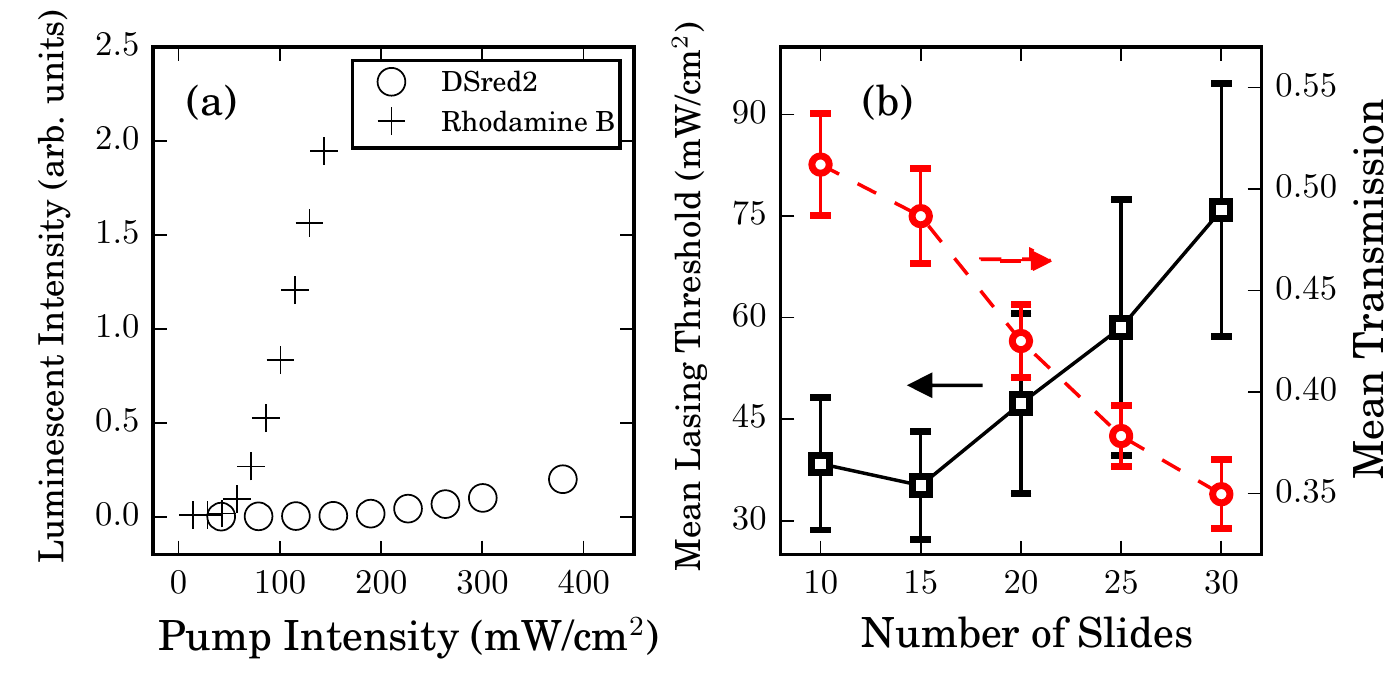}
   \caption{(a) Luminescent output vs. pump intensity under femtosecond pumping for both DsRed2 ($\circ$) and Rhodamine B (+) at a single lasing region. Each point is the average of ten single-shot spectra. (b) Average lasing threshold of 76 lasing regions for Rhodamine B under femtosecond pumping vs. number of layers in the input half of the random stack (the ``input coupler'') compared to the mean transmission of the input coupler. It is clear that the increase in measured lasing threshold results from the increasing rejection of pump energy with the greater number of layers.}
   \label{Figure3}
\end{figure}

In addition to the spectral line narrowing, laser emission is typically characterized by a threshold-like power behavior, shown in Figure \ref{Figure3} for the case of the femtosecond pumping. Moving the stack transversely with respect to the pump beam exposed different random cavity configurations due to slight variations in the glass and air gap thickness, which resulted in lasing spectra with differing emission peaks within the gain bandwidth. Figure \ref{Figure3} shows the spectrally integrated emission signal from a single lasing area of the stack using either the DsRed2 solution ($\circ$) or a solution of Rhodamine B (+) of similar optical density. An abrupt increase in the output power is clearly seen at pump intensity of 150~mW/cm$^{2}$ in DsRed2 and at 50~mW/cm$^{2}$ in the Rhodamine solution. Each medium exhibits an approximately linear power growth above threshold.

Pumping the random laser with nanosecond or femtosecond pulses resulted in strikingly different lasing thresholds. To compare the two cases, we collected the threshold values from 20 distinct locations across the random stack. For a stack of 60 layers, ultrafast femtosecond pumping initiated lasing in DsRed2 at average pump intensities of 0.230~W/cm$^{2}$, about 2 orders of magnitude lower than its nanosecond counterpart, which required pulse intensities of order 12.5~W/cm$^{2}$. We note that the nanosecond pump intensity necessary to elicit lasing well exceeds the $\sim$1~W/cm$^{2}$ shown to cause strong photobleaching in DsRed2 \cite{Drobizhev2012} and as such it was not possible to find a region of the stack that would lase under continuous nanosecond pumping. While the average femtosecond threshold of 0.230~W/cm$^{2}$ is well above the 0.03~W/cm$^{2}$ known to allow steady state fluorescent output, singular lasing regions with thresholds on the order of 0.100~W/cm$^{2}$ were not uncommon. When these areas were pumped above threshold the lasing output declined rapidly and fell below 25\% of the initial intensity after 500 pulses (25 s at 20 Hz). Lasing was still evident after 5000 pulses although the output at that point had reduced to 5-10\% of the initial intensity.

To determine the nature of the discrepancy in lasing threshold between nano- and femtosecond pumping, the lasing behaviour of this 1D scattering system was checked using Rhodamine B as the gain medium as it was capable of sustained lasing under nanosecond excitation. The lasing threshold and input transmittance for each pump source was determined at 76 distinct locations across the aperture of the assembly for stacks comprised of 20, 30, 40, 50, and 60 total glass layers. The results for femtosecond pumping are shown in Figure \ref{Figure3}(b). Average lasing thresholds for ultrafast pumping ranged from 0.030 to 0.075~W/cm$^{2}$ while the average threshold for nanosecond excitation varied between 7.1 and 12.2~W/cm$^{2}$ matching the 2 order of magnitude difference observed in the DsRed2 gain medium. Because these two fluorophores have roughly similar photo-physical characteristics, the observed threshold difference is unlikely to be particular to these media.

\begin{figure}[tb]
   \centering
   \includegraphics[width=.85\columnwidth]{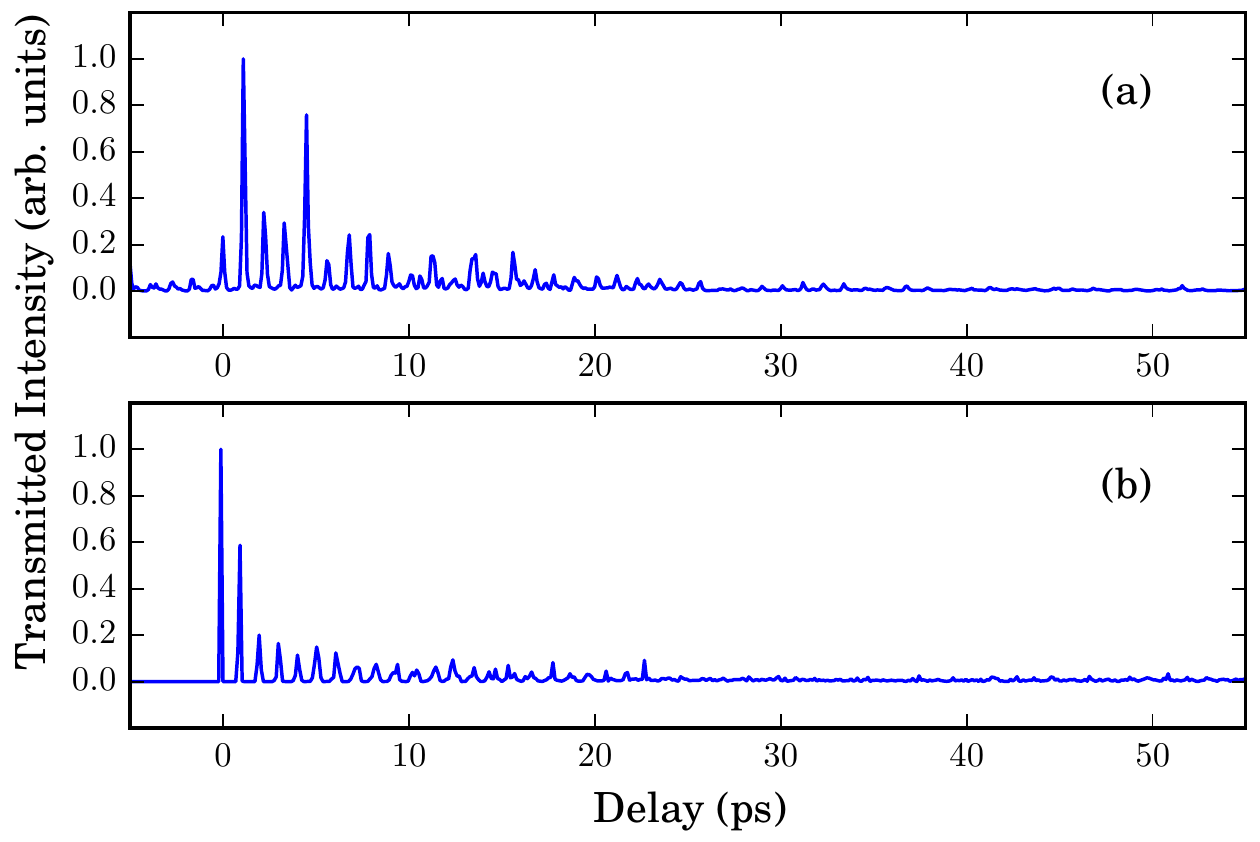}
\caption{Pulse train generated by the femtosecond pump pulse traversing the coverslip stack and cuvette assembly, (a) measured by optical cross correlation and (b) calculated by means of the transfer matrix analysis for a 1D system with a similar layer number, thickness, and refractive index.}
\label{Figure4}
\end{figure}

To explain the difference in the lasing thresholds corresponding to the two pumping regimes, we note the following separation of the time scales:
\begin{equation}\label{Timescales}
    \tau_\text{fs} \ll \tau_\text{c} \ll \tau_\text{sp} \ll \tau_\text{ns},
\end{equation}
where $\tau_\text{fs} \approx 50$ fs and  $\tau_\text{ns} \approx 300$ ns are the durations of the femto- and nanosecond pump pulses, respectively;  $\tau_\text{sp} \approx 3.6$ ns or 2.28 ns is the spontaneous lifetime of DsRed \cite{Lounis2001} or Rhodamine B \cite{Sauer2010} respectively, and  $\tau_\text{c} \approx 30$ ps is the lifetime of an average cavity mode. The latter was determined by performing the time resolved measurement of the bare cavity decay (without gain medium), after it has been excited by an ultrashort pump pulse. Typical cavity decay, obtained by optical cross-correlation of the transmitted light with a reference pulse (see Methods), is plotted in Fig.\ref{Figure4}(a). As expected for a layered 1D medium (the results of numerical calculations are plotted in Fig.\ref{Figure4}(b)), it shows a random pulse train with an average time between pulses corresponding to the round-trip time in a single layer.

Because of the outlined time scale separation (\ref{Timescales}), a femtosecond pump pulse will deliver its energy before either the lasing process or the spontaneous decay has time to deplete the excited state population. Using the ``four-level laser'' model \cite{MilonniEberlyBook}, one finds for the population inversion $\Delta N$ on the lasing transition:
\begin{equation}\label{fsInversion}
    \Delta N_\text{fs} \approx R_\text{fs} \tau _\text{fs},
\end{equation}
where $R_\text{fs}=I_\text{fs}\sigma _\text{p}/\hbar \omega _\text{p}$ is the pumping rate expressed through the femtosecond pump intensity ($I_\text{fs}$) and absorption cross-section ($\sigma _\text{p}$) at the pump frequency ($\omega _\text{p}$). To arrive at this simple expression, we used the fact that the absorption of the pump beam is high (measured optical density of the Rhodamine solution, OD$>$1) and far from saturation. Equation (\ref{fsInversion}) supports the observed dependence of the lasing threshold on the number of layers in the random cavity (Fig.\ref{Figure3}(b)). Indeed, as the number of layers increases from 20 to 60, the amount of pump energy delivered to the gain medium (and hence, the pumping rate $R$) falls off due to the decreasing spectral width of random transmission modes, leading to the correspondingly lower population inversion and higher lasing thresholds.

In the case of the nanosecond pumping, inequalities (\ref{Timescales}) justify invoking the steady-state solution of the laser rate equations, which results in the following population inversion:
\begin{equation}\label{nsInversion}
    \Delta N_\text{ns} \approx R_\text{ns} \tau _\text{sp} = R_\text{ns} \tau _\text{ns} \times \frac{\tau _\text{sp}}{\tau _\text{ns}}.
\end{equation}

Comparing expressions (\ref{fsInversion}) and (\ref{nsInversion}), we note that given the same total pulse energies, $R_\text{fs} \tau _\text{fs} = R_\text{ns} \tau _\text{ns}$, the ratio between the population inversion induced by femto- and nanosecond pump pulses is $\tau _\text{ns}/\tau _\text{sp} \approx 100$, explaining the observed two orders of magnitude difference between the corresponding lasing thresholds. To check this conclusion, we varied $\tau _\text{sp}$ by replacing the protein medium with two different solutions of the laser dye LDS 750.

\begin{figure}[tb]
\centering
   \includegraphics[width=.85\columnwidth]{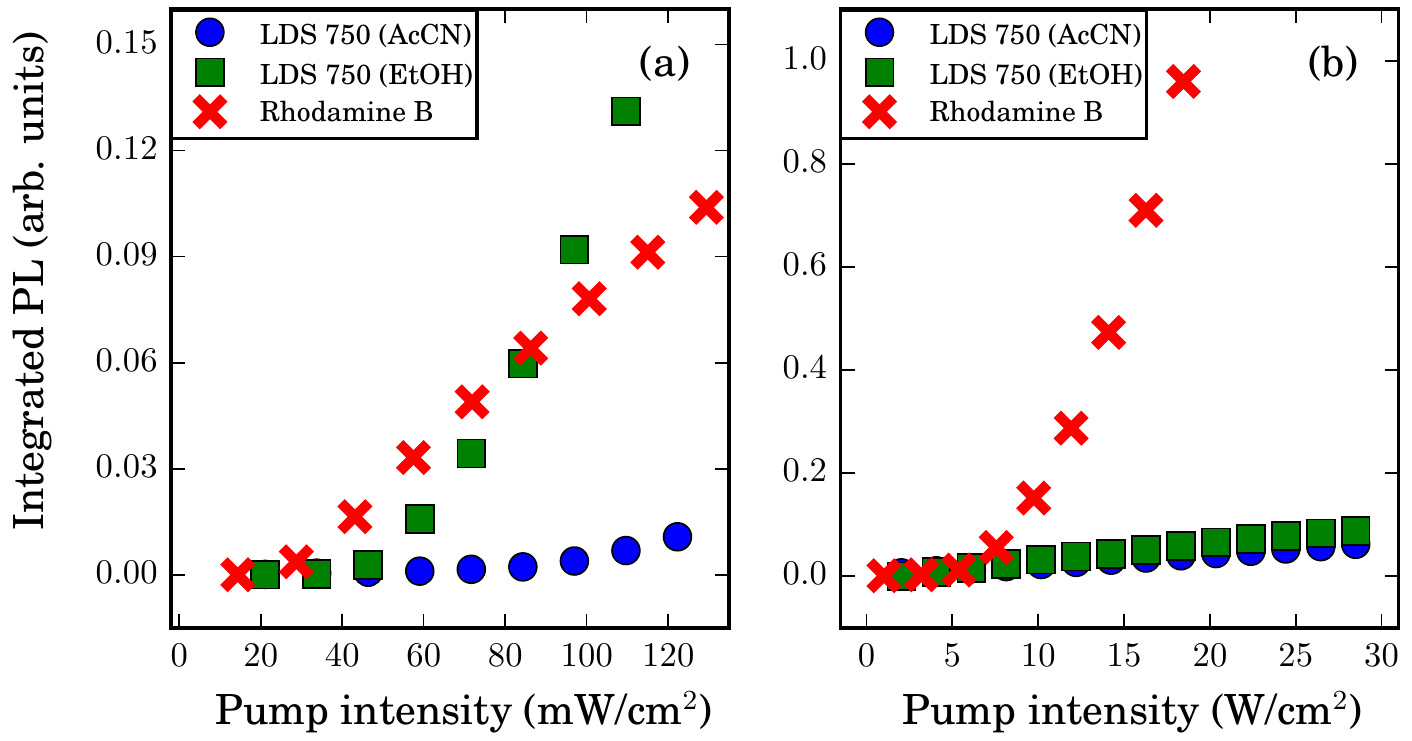}
\caption{Mean photoluminescent (PL) output from a 30 layer stack containing the short $\tau_{sp}$ dye LDS 750 in acetonitrile and ethanol pumped via femtosecond (a) or nanosecond (b) pulses. The 30 layer average outputs for Rhodamine B are provided for comparison. }
\label{Figure5}
\end{figure}

The fluorescence lifetime of LDS 750 is dependent on solvent viscosity, $\eta$, according to relation
\begin{equation}\label{solventTau}
\frac{1}{\tau_{sp}}=1.06 + \frac{1.90}{\eta},
\end{equation}
with $\eta$ expressed in cP \cite{Castner1987}.
In acetonitrile (AcCN, $\eta=0.36$), $\tau_{sp}$ is known to be 160 ps while in ethanol (EtOH, $\eta=1.24$) Eq.\ref{solventTau} predicts a value of $\tau_{sp}=360$ ps, i.e. shorter by an order of magnitude than the lifetime of DsRed2 or Rhodamine B. A 30 layer stack containing LDS 750 in either AcCN or EtOH was scanned for threshold behaviour at 53 distinct points in the manner similar to the DsRed2 and Rhodamine B media. The mean output is shown in Figure \ref{Figure5} along with the data for Rhodamine B for comparison. The shortest lifetime medium (Fig.\ref{Figure5}(a), AcCN, blue circles) showed consistent lasing under femtosecond pumping with an average threshold pump intensity of 0.090~W/cm$^{2}$ as did the LDS 750 in ethanol (Fig.\ref{Figure5}(a), green squares) with a mean threshold of 0.035~W/cm$^{2}$.

For nanosecond excitation, reducing the $\tau_{sp}$ of the medium suppressed all consistent lasing output up to at least a pump intensity of 30~W/cm$^{2}$, well above the nanosecond threshold observed in the longer lifetime media. The increase of the ratio $\tau_\text{sp}/\tau_\text{ns}$ to $\sim1000$ in these shorter fluorescence lifetime gain media suggests that nanosecond pumped lasing could be achieved at 2-3 times this intensity, however nanosecond intensity greater than 30~W/cm$^{2}$ exceeded the damage threshold of some components of the experimental apparatus and the threshold search stopped there.

\section{Summary}
To summarize, we have demonstrated lasing of the fluorescent protein DsRed2 in a 1D random cavity. We showed that optically pumping the medium with femtosecond, rather than nanosecond, pulses reduces the required average pump intensity by more than two orders of magnitude. Ultrafast pumping may therefore be preferred when searching for random lasing in biological systems with low damage threshold or media with lower scattering efficiency. Our successful demonstration of low-threshold lasing suggests that ultrafast pumped DsRed may be a suitable amplifying material in more complex scattering media. However, the random lasing output from DsRed when pumped well above threshold was found to decay on the scale of a few tens of seconds. This decay is in contrast to the robustness observed in eGFP, which is reported to have survived over 5,000 excitations at 200 times the lasing threshold without any noticeable loss of output intensity\cite{Gather2011}. The signal loss seen here with DsRed likely arises from the overall higher pump intensities required to reach the lasing threshold in a relatively low Q factor of the 1D scattering medium as the photobleaching yield of DsRed has not been found to differ greatly from that of eGFP \cite{Gross2000, Lounis2001, Shaner2008}. It is worth noting that several engineered derivatives of DsRed, such as Tag-RFP-T or mCherry  have better photobleaching resistance \cite{Shaner2004, Shaner2008} and may also prove suitable as gain media in low finesse resonators for potential bio-photonic applications.

\section{Methods}
The gene coding for DsRed was amplified from the plasmid  pDsRed2-N1 (Clontech) as published\cite{Bach2008}. The amplified gene was subcloned into pMal (NEB) and expressed as a fused protein to the C' terminus of maltose-binding protein. Recombinant DsRed protein was produced in Escherichia coli strain BL-21. A starter culture was commenced by picking a single colony into Luria-Bertani broth supplemented with 1\% glucose and 50 ug/mL ampicillin at 37~\textdegree C overnight. The next day, the culture was diluted 1:100 in the same medium and incubated at 37~\textdegree C until an optical density of 0.6 at 600 nm was reached. Cells were induced with 0.4 mM IPTG at room temperature for 16 hours. Cells were then harvested at 8,000 rpm for 15 minutes, and resuspended in binding buffer (20 mM Tris-HCl, 1 mM EDTA, 5\% glycerol, pH 7.2). DsRed protein was purified from the induced cells after sonication and centrifugation at 15,000 rpm for 30 min at 4~\textdegree C using an amylose resin (NEB) and according to the manufacturer's instructions. Purified protein was dialyzed overnight in Phosphate Buffered Saline (PBS) overnight at 4~\textdegree C using dialysis bag  (cut-off 8,000). A concentration of 38 mg/mL of DsRed protein was calculated using the Bradford assay and using bovine serum albumin as control.

The time domain behaviour of the 1-D scatterer configured for lasing was measured by generating the sum frequency of a 35 fs reference pulse centered at 800~nm and the 530 nm femtosecond pump beam transmitted through the random stack containing an ethanol blank.
The delay of the reference beam was was scanned with a variable delay line, while the cross correlation signal between 325-350 nm was recorded. The results showed the initial pump pulse broken into a random pulse train which decayed on a time scale of up to 40 ps as shown in Fig.\ref{Figure4}(a). Numerical simulations of the time-dependent transmission through a stack of glass layers with a similarly distributed random thickness shows qualitatively similar behavior (Fig.\ref{Figure4}(b)).

The authors would like to thank Dr. A. A. Milner for his collaboration at the outset of this work.

\end{document}